\documentclass[twocolumn]{aastex631}  
\usepackage[mathlines]{lineno}  

\shorttitle{Masses for Luminous Giants}
\shortauthors{K. Cao \& Pinsonneault}
\usepackage{CJKutf8}

\usepackage{graphicx}	
\usepackage{amsmath}	
\usepackage{amssymb}	
\allowdisplaybreaks

\newcommand{\NewEdit}[1]{#1} 

\begin{document}

\title{Towards Accurate Asteroseismic Masses for Luminous Giants}

\correspondingauthor{Kaili Cao}
\email{cao.1191@osu.edu}

\author[0000-0002-1699-6944]{Kaili Cao (\begin{CJK*}{UTF8}{gbsn}曹开力\end{CJK*}$\!\!$)}
\affiliation{Center for Cosmology and AstroParticle Physics (CCAPP), The Ohio State University, 191 West Woodruff Ave, Columbus, OH 43210, USA}
\affiliation{Department of Physics, The Ohio State University, 191 West Woodruff Ave, Columbus, OH 43210, USA}

\author[0000-0002-7549-7766]{Marc H. Pinsonneault}
\affiliation{Center for Cosmology and AstroParticle Physics (CCAPP), The Ohio State University, 191 West Woodruff Ave, Columbus, OH 43210, USA}
\affiliation{Department of Astronomy, The Ohio State University, 140 West 18th Avenue, Columbus, OH 43210, USA}

\begin{abstract}

Asteroseismology, the study of stellar oscillations, provides high-precision measurements of masses and ages for red giants. Scaling relations are a powerful tool for measuring fundamental stellar parameters, and the derived radii are in good agreement with fundamental data for low-luminosity giants. However, for luminous red giant branch (RGB) stars, there are clear systematic offsets. In APOKASC-3, the third joint spectroscopic and asteroseismic catalog for evolved stars in the Kepler fields, we tied \NewEdit{asteroseismic} radii to a reference system based on Gaia astrometry by introducing correction factors. This work proposes an alternative formulation of the correction scheme, which substantially reduces the sensitivity of the results to the technique used to infer mean density from frequency spacings. Compared to APOKASC-3, our adjusted correction scheme also reduces fractional discrepancies in median masses and ages of lower RGB and upper RGB within the $\alpha$-rich population from $6.65\%$ to $1.72\%$ and from $-21.81\%$ to $-9.55\%$, respectively. For the $\alpha$-poor population, the corrected mass scale leads to an improved agreement between theory and observation of the surface carbon-to-nitrogen abundance ratio, a significant diagnostic of the first dredge-up.

\end{abstract}


\section{Introduction} \label{sec:intro}

Time domain data from space missions has had an enormous impact on stellar astrophysics. One of the most dramatic examples is the field of asteroseismology, or the study of stellar oscillations. Virtually all evolved red giant stars are observed as high-amplitude oscillators in these missions. When combined with spectroscopy, the oscillation patterns can be used to infer important properties such as mass, radius, and age \citep[see][for a review]{Chaplin2013ARA&A}. Solar-like oscillations are typically characterized by two figures of merit, the frequency of the maximum power $\nu_{\max}$ and the large frequency spacing $\Delta\nu$ \citep{Gough1986HMPS, Kjeldsen1995A&A, Miglio2021A&A}. Measurements of stellar properties are then based on theoretical scaling relations of these two quantities.

\NewEdit{Over the last $15$ years, sophisticated techniques for measuring these global properties have been developed; see the appendices of \citet{Pinsonneault2025ApJS} for a summary of eleven commonly used methods. All four major exoplanet transit missions---CoRoT \citep{Baglin2003AdSpR}, Kepler \citep{Borucki2010Sci, Koch2010ApJ}, K2 \citep{Howell2014PASP}, and TESS \citep{Ricker2015JATIS}---have published asteroseismic data. However, the Kepler mission has played a central role in population asteroseismology, in large part because of its exceptionally long dwell time and high data quality.}

The APOKASC project \citep{Pinsonneault2014ApJS, Pinsonneault2018ApJS, Pinsonneault2025ApJS} was designed to translate these asteroseismic data into precise and accurate stellar masses, radii and ages. Three key ingredients were identified. First, multiple analysis methods were used to identify outliers and to make more precise measurements. Second, theoretical stellar interiors models were used to map the observed $\Delta\nu$ onto the mean density \citep{Pinsonneault2018ApJS, Pinsonneault2025ApJS}; this corrects for the fact that the observed frequencies are not in the asymptotic regime, and therefore the real spacings are not uniform \citep{White2011ApJ, Mosser2012A&A, Stello2014ApJ}. This step requires computing a theoretical spectrum and mapping it onto the observed one with a weighting scheme. Finally, the measurements were calibrated to a fundamental scale. In APOKASC-2 \citep{Pinsonneault2018ApJS}, the zero point of the $\nu_{\max}$ relationship was set to be consistent with masses of members of open star clusters. While this empirical correction has a solid foundation \citep{Zinn2019ApJ}, offsets between asteroseismic and fundamental radii were seen in luminous stars.

APOKASC-3 \citep{Pinsonneault2025ApJS}, the final version of the joint catalog of evolved stars with spectroscopic data from APOGEE \citep{Majewski2010IAUS, Majewski2017AJ, Ahumada2020ApJS, Abdurrouf2022ApJS} and asteroseismic data from Kepler \citep{Gilliland2010PASP, Kjeldsen2010AN}, is a landmark dataset in these efforts. In APOKASC-3, asteroseismic radii were compared with radii inferred from a combination of luminosity \citep[from Gaia and extinction map data;][]{Gaia2016A&A, Gaia2023A&A} and spectroscopic $T_{\rm eff}$ (from the APOGEE survey). That paper applied a $\nu_{\max}$-dependent correction term to align the two scales, which corrects for the $\nu_{\max}$ dependent trends seen for luminous giants.  However, because the scaling relationships between radius and mass have different functional forms, it is not a priori obvious that concordance with the empirical radius scale places the masses on a consistent system, and there is evidence that the adopted procedure did induce some systematics in the relative masses of high\NewEdit{-} and low\NewEdit{-}luminosity stars \citep{Ash2025ApJ}.

Motivated by this, we propose and investigate an alternative formulation of the correction. As we will demonstrate, it yields some improvements and serves as a measure of systematics. This letter is structured as follows. In Section~\ref{sec:meth}, we introduce the formalism of different correction schemes and describe how we implement each of them on APOKASC-3 stars. Then we present our results in Section~\ref{sec:res} and discuss their implications in Section~\ref{sec:summ}.

\section{Methods} \label{sec:meth}

For solar-like oscillators, the asteroseismic scaling relations for mass and radius can be written as
\begin{equation}
    \frac{M_{\rm seis}}{{\rm M}_{\odot}} = f_{\nu_{\rm max}}^3 \left( \frac{\nu_{\rm max}}{\nu_{{\rm max}, \odot}} \right)^3 \left( \frac{f_{\Delta\nu} \Delta \nu}{\Delta \nu_{\odot}} \right)^{-4} \left( \frac{T_{\rm eff}}{T_{{\rm eff}, \odot}} \right)^{1.5}
    \label{eq:scalingM}
\end{equation}
and
\begin{equation}
    \frac{R_{\rm seis}}{{\rm R}_{\odot}} = f_{\nu_{\rm max}} \left( \frac{\nu_{\rm max}}{\nu_{{\rm max}, \odot}} \right) \left( \frac{f_{\Delta\nu} \Delta \nu}{\Delta \nu_{\odot}} \right)^{-2} \left( \frac{T_{\rm eff}}{T_{{\rm eff}, \odot}} \right)^{0.5},
    \label{eq:scalingR}
\end{equation}
where $\nu_{{\rm max}, \odot} = 3076 \,{\rm \mu Hz}$ and $\Delta \nu_{\odot} = 135.1416 \,{\rm \mu Hz}$ are reference solar values \NewEdit{from APOKASC-2 \citep{Pinsonneault2018ApJS}}, $f_{\nu_{\rm max}}$ and $f_{\Delta\nu}$ denote correction factors; $T_{\rm eff}$ is effective temperature from spectroscopic data, and $T_{{\rm eff}, \odot} = 5772 \,{\rm K}$ \NewEdit{\citep{Mamajek2015arXiv}}.

The frequency spacing correction factor, $f_{\Delta\nu} = \langle \rho \rangle^{1/2} / \Delta\nu$, accounts for the nonlinear mapping between $\Delta\nu$ and the mean density $\langle \rho \rangle$. \NewEdit{The theoretically predicted frequency spacings, as well as those seen in the data, are not strictly uniform. As a result, a weighting scheme must be used to collapse the theoretical spectrum into a single value.} In APOKASC-3 \citep{Pinsonneault2025ApJS}, recommended $f_{\Delta\nu}$ values were inferred from GARSTEC \citep{Weiss2008Ap&SS} models and the \citet{Mosser2012A&A} weighting approach\NewEdit{, which uses an empirical method for setting the width over which the frequency spacings are used}. Since $f_{\Delta\nu}$ \NewEdit{is a function of mass, metallicity, and radius}, an iterative process was used to account for the mass dependence. Unlike $f_{\Delta\nu}$, $f_{\nu_{\rm max}}$ was just a function of $\nu_{\rm max}$ (in ${\rm \mu Hz}$)
\begin{equation}
    f = [1 + a (\ln \nu_{\rm max})^3 + b (\ln \nu_{\rm max})^2 + c (\ln \nu_{\rm max}) + d]^{-1},
    \label{eq:poly_fit}
\end{equation}
with fitting coefficients $a$, $b$, $c$, and $d$; $f$ is kept fixed for $\nu_{\rm max} > 50 \,{\rm \mu Hz}$. Note that we have combined Eqs.~(4) and (5) in \citet{Pinsonneault2025ApJS} and omitted the subscript as this functional form will be used for other purposes in Section~\ref{sec:res}. The APOKASC-3 coefficients were empirically determined by setting
\begin{equation}
    \left\langle f_{\nu_{\rm max}} \left( \frac{\nu_{\rm max}}{\nu_{{\rm max}, \odot}} \right) \left( \frac{f_{\Delta\nu} \Delta \nu}{\Delta \nu_{\odot}} \right)^{-2} \left( \frac{T_{\rm eff}}{T_{{\rm eff}, \odot}} \right)^{0.5} \right\rangle = \frac{\langle R_{\rm Gaia} \rangle}{{\rm R}_{\odot}},
    \label{eq:gaia_fnumax}
\end{equation}
where $R_{\rm Gaia}$ is stellar radius derived from Gaia \NewEdit{parallax}, photometry, and spectroscopic temperature.

APOKASC-3 also provided two alternative sets of $f_{\Delta\nu}$, $f_{\nu_{\rm max}}$, $M_{\rm seis}$, and $R_{\rm seis}$ values. Both sets used the \citet{White2011ApJ} weighting approach\NewEdit{, which uses a weighted linear fit to $\Delta\nu$ as a function of frequency, as opposed to the Mosser method, which uses a weighted average across an empirically defined frequency window. One} of them also used GARSTEC models, while the other used models from \citet{Sharma2016ApJ}. The $\nu_{\rm max}$ correction procedure is the same as GARSTEC+Mosser. These alternative sets serve as a measure of systematics, which we will further explore in Section~\ref{sec:res}.

While Eq.~(\ref{eq:gaia_fnumax}) is an entirely reasonable correction scheme, APOKASC-3 did hypothesize that $f_{\Delta\nu}$ values from models were accurate. However, \citet{Ash2025ApJ} indicated that neither $f_{\Delta\nu}$ nor the radius calibration produced exact agreement between asteroseismic measurements and the Gaia-based fundamental reference system using $\alpha$-rich giants as a diagnostic tool. This old population has a well-defined characteristic age, corresponding to a well-defined peak in mass. On the lower RGB, the mean mass is independent of $\nu_{\rm max}$, consistent with astrophysical expectations. However, for luminous giants, the median mass in APOKASC-3 becomes a function of $\nu_{\rm max}$\NewEdit{, with higher mass on the upper RGB}. This is opposite to the effect induced by mass loss, and it represents an indication of a systematic mass error \citep[see][for a fuller discussion]{Ash2025ApJ}. Resting on this observation, here we propose an alternative formulation, which considers the $f$ obtained in Eq.~(\ref{eq:gaia_fnumax}) as a correction factor for radius, not for $\nu_{\rm max}$, i.e.,
\begin{equation}
    \left\langle f_{R} \left( \frac{\nu_{\rm max}}{\nu_{{\rm max}, \odot}} \right) \left( \frac{f_{\Delta\nu} \Delta \nu}{\Delta \nu_{\odot}} \right)^{-2} \left( \frac{T_{\rm eff}}{T_{{\rm eff}, \odot}} \right)^{0.5} \right\rangle = \frac{\langle R_{\rm Gaia} \rangle}{{\rm R}_{\odot}}.
    \label{eq:gaia_fR}
\end{equation}
Numerically, $f_{R}$ has the same values as $f_{\nu_{\rm max}}$; but meanwhile, it has a different meaning\NewEdit{: $f_{\nu_{\rm max}}$ is solely tied to $\nu_{\rm max}$, while $f_{R}$ is tied to $R_{\rm seis}$, which is affected by both $\nu_{\rm max}$ and $\Delta\nu$. Since the degeneracy between corrections to $\nu_{\rm max}$ and $\Delta\nu$ captured by $f_{R}$ cannot be broken with given information, and systematics involved in seismic gravity are usually much smaller than uncertainties of spectroscopic gravity, it is reasonable to assume that $g_{\rm seis}$ is unaffected by our new correction scheme. Therefore, we have $M_{\rm seis} \propto R_{\rm seis}^2$, and $f_{R}$} leads to a different mass scaling relation
\begin{equation}
    \frac{M_{\rm seis}}{{\rm M}_{\odot}} = f_R^2 \left( \frac{\nu_{\rm max}}{\nu_{{\rm max}, \odot}} \right)^3 \left( \frac{f_{\Delta\nu} \Delta \nu}{\Delta \nu_{\odot}} \right)^{-4} \left( \frac{T_{\rm eff}}{T_{{\rm eff}, \odot}} \right)^{1.5}.
    \label{eq:scalingM_fR}
\end{equation}

There are two ways of implementing this $f_R$ correction scheme. Comparing Eqs.~(\ref{eq:scalingM}) and (\ref{eq:scalingM_fR}), the ``raw'' way is to simply divide tabulated $M_{\rm seis}$ values by the corresponding $f_R$ values, i.e., to convert $f_{\nu_{\rm max}}^3$ to $f_R^2$. However, as $M_{\rm seis}$ changes, $f_{\Delta\nu}$ also changes, and a new iterative process is needed to account for this feedback. Specifically, for each star, we start with the initial guess of tabulated $M_{\rm seis} / f_{\nu_{\rm max}}$, use it to obtain an updated $f_{\Delta\nu}$ (see Section~\ref{sec:res} for details), and compute a new $M_{\rm seis}$ according to Eq.~(\ref{eq:scalingM_fR}); we repeat this process until the absolute value of the fractional change in $M_{\rm seis}$ is below $\epsilon = 10^{-12}$. We refer to the implementation which adjusts $f_{\Delta\nu}$ as the adjusted way.

To investigate the systematics in asteroseismic masses, we conduct two sets of tests using APOKASC-3 stars. Specifically, we select red giant branch (RGB) stars with $0.8 \,{\rm M}_{\odot} < M_{\rm seis} < 2.0 \,{\rm M}_{\odot}$ (using recommended $M_{\rm seis}$ values) and $-0.45 < [{\rm Fe}/{\rm H}] < +0.45$ from both ``Gold'' and ``Silver'' samples --- the former has better quality, but the latter contains a larger fraction of luminous giants. Thanks to reliable evolutionary state determination \citep{Elsworth2019MNRAS, Vrard2024A&A}, we believe that this combined sample has a low contamination fraction from asymptotic giant branch (AGB) stars that could have experienced significant mass loss.\footnote{We select targets with {\tt \textquotesingle Evol\_State\textquotesingle} of either {\tt \textquotesingle RGB\textquotesingle} ($4727$ stars) or {\tt \textquotesingle RGB\_AGB\textquotesingle} ($809$ stars), as {\tt \textquotesingle RGB\_AGB\textquotesingle} stars are very likely RGB stars.} To further mitigate the impact of mass loss, we use median statistics instead of mean statistics.

Among these $5536$ RGB stars, there are $1138$ $\alpha$-rich ones and $4398$ $\alpha$-poor ones. \NewEdit{For a better understanding of the parameter space covered by the data, we refer the readers to \citet{Pinsonneault2025ApJS}: Fig.~2 for $[{\rm Fe}/{\rm H}]$, Fig.~4 for $\nu_{\rm max}$, Fig.~17 for $M_{\rm seis}$ and $R_{\rm seis}$, etc.} Following \citet{Ash2025ApJ}, our first set of tests check the consistency of masses and ages between lower RGB and upper RGB within the $\alpha$-rich population \citep{Bensby2003A&A}, which is assumed to be a collection of stars within the Milky Way with similar mass and age. The second set of tests study theoretical implications for the $\alpha$-poor population. The results are presented in the next section.

\section{Results} \label{sec:res}

\begin{figure}
    \centering
    \includegraphics[width=\columnwidth]{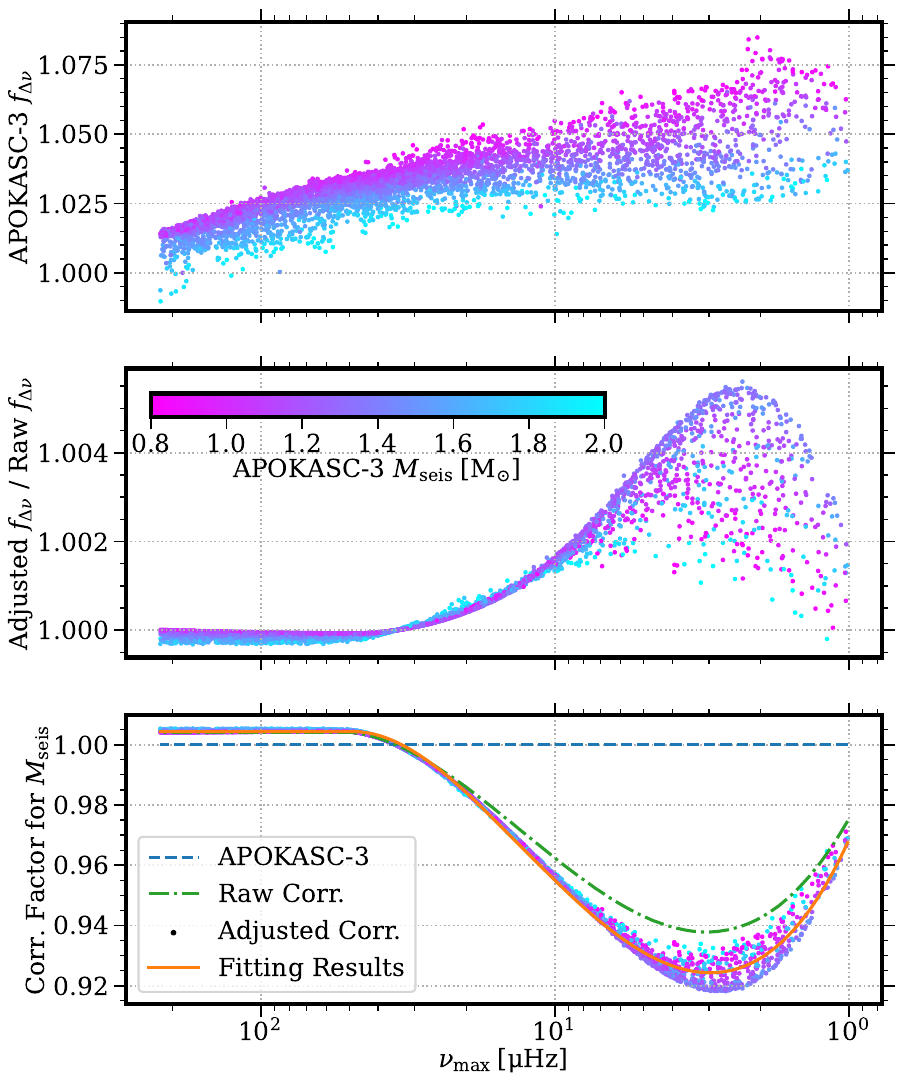}
    \caption{\label{fig:fdnu_fm_numax}A comparison between different correction schemes for asteroseismic masses. All quantities are plotted as functions of $\nu_{\rm max}$, with the lower RGB shown on the left and the upper RGB shown on the right. The upper panel shows the published $f_{\Delta\nu}$ from APOKASC-3 for our sample. In the middle panel, we plot the ratio between the adjusted (see text) $f_{\Delta\nu}$ and the raw $f_{\Delta\nu}$. In the lower panel, we illustrate the net effect on the masses relative to the published APOKASC-3 values, which correspond to unity and are shown by the blue dashed line. The raw $f_R$ correction, holding $f_{\Delta\nu}$ fixed \NewEdit{(so that the mass correction factor is simply $1 / f_R$)}, is shown by the green dash-dotted line. The adjusted $f_R$ correction, including feedback in the $f_{\Delta\nu}$ calculation, is shown by colored dots. Although it depends on both $\nu_{\rm max}$ and $M_{\rm seis}$, simple fitting results \NewEdit{using Eq.~(\ref{eq:poly_fit}) (}ignoring mass dependence\NewEdit{)} are also shown by the orange solid curve. The fitting coefficients are tabulated in Table~\ref{tab:poly_coeffs}.}
\end{figure}

Fig.~\ref{fig:fdnu_fm_numax} compares different correction schemes for asteroseismic masses. From the upper panel, it is clear that APOKASC-3 $f_{\Delta\nu}$ depends on both $\nu_{\rm max}$ and $M_{\rm seis}$. A closer look would reveal that, at each $\nu_{\rm max}$, $f_{\Delta\nu}$ decreases as $M_{\rm seis}$ increases. Since $M_{\rm seis}$ scales as \NewEdit{$f_{\Delta\nu}^{-4}$}, in our adjusted $f_R$ correction scheme, this negative correlation amplifies the additional mass correction induced by switching from $f_{\nu_{\rm max}}^3$ to $f_R^2$. To capture $f_{\Delta\nu}$ trends in APOKASC-3, we train a {\sc scikit-learn} nu-support vector regression (nu-SVR), which \NewEdit{uses observationally motivated features $\nu_{\rm max}$, mass, and corrected metallicity $[{\rm Fe}/{\rm H}]_{\rm corr} = [{\rm Fe}/{\rm H}] + 0.625 [\alpha/{\rm Fe}]$ as input, and} yields an average score of $0.9697$ in a 5-fold cross validation.\footnote{This refers to the {\tt test\_score} component of the return value of {\tt sklearn.model\_selection.cross\_validate}.} The score is close to $1$, indicating that our nu-SVR can well replicate the dependence of $f_{\Delta\nu}$ on $\nu_{\rm max}$, mass, and $[{\rm Fe}/{\rm H}]_{\rm corr}$. For each star in our sample, we start from APOKASC-3 $M_{\rm seis} / f_{\nu_{\rm max}}$ and use an iterative process (outlined in Section~\ref{sec:meth}) to find the ``adjusted'' $f_{\Delta\nu}$ and $M_{\rm seis}$; the resulting changes in $f_{\Delta\nu}$ are shown in the middle panel.

\begin{table}
    \caption{\label{tab:poly_coeffs}Fitting coefficients for correction factors for asteroseismic masses and ages. The fitting function is Eq.~(\ref{eq:poly_fit}), which only depends on $\nu_{\rm max}$ and ignores mass dependence (see the lower panel of Fig.~\ref{fig:fdnu_fm_numax}). The correction factors are supposed to be applied to masses and ages in the APOKASC-3 main table \citep{Pinsonneault2025ApJS}. Fitting results for both raw and adjusted $f_R$ corrections are included.}
    \centering
    \begin{tabular}{lcccc}
    \hline
        Property (Scheme) & $a$ & $b$ & $c$ & $d$ \\
    \hline
        Mass (Raw) & $0.0061$ & $-0.0465$ & $0.0809$ & $0.0251$ \\
        Mass (Adjusted) & $0.0079$ & $-0.0590$ & $0.1003$ & $0.0324$ \\
    \hline
        Age (Raw) & $-0.0166$ & $0.1288$ & $-0.2282$ & $-0.0741$ \\
        Age (Adjusted) & $-0.0211$ & $0.1604$ & $-0.2779$ & $-0.0922$ \\
    \hline
    \end{tabular}
\end{table}

Finally in the lower panel, we compare the mass correction factors according to the three different correction schemes. For lower RGB stars ($\nu_{\rm max} < 23.50 \,{\rm \mu Hz}$), there is no change ($< 1\%$ in mass) with either raw or adjusted correction. For upper RGB stars, the raw correction scheme lowers $M_{\rm seis}$ by up to $6.22\%$ (at $\nu_{\rm max} = 2.98 \,{\rm \mu Hz}$), while the adjusted correction scheme reduces $M_{\rm seis}$ by up to $8.18\%$ (at $\nu_{\rm max} = 2.63 \,{\rm \mu Hz}$). It is an interesting feature that the $f_{\Delta\nu}$ feedback also significantly contributes to the mass correction; the fitting results (green dash-dotted and orange solid curves) indicate that its contribution is roughly proportional to that of $f_R$. The fitting coefficients are tabulated in Table~\ref{tab:poly_coeffs}.

This table also includes correction factors for stellar ages reported in APOKASC-3, which come from a similar nu-SVR trained on APOKASC-3 data (average score in cross validation: $0.9997$). \NewEdit{We refer interested readers to Section~3.3.2 of \citet{Pinsonneault2025ApJS} for age determination in APOKASC-3. The raw and corrected ages are derived by feeding the raw and corrected masses (along with $\nu_{\rm max}$ and $[{\rm Fe}/{\rm H}]_{\rm corr}$) to the nu-SVR, respectively. This nu-SVR is also used for stellar ages in the no $\nu_{\rm max}$ case (see below).}

\begin{figure*}
    \centering
    \includegraphics[width=0.9\textwidth]{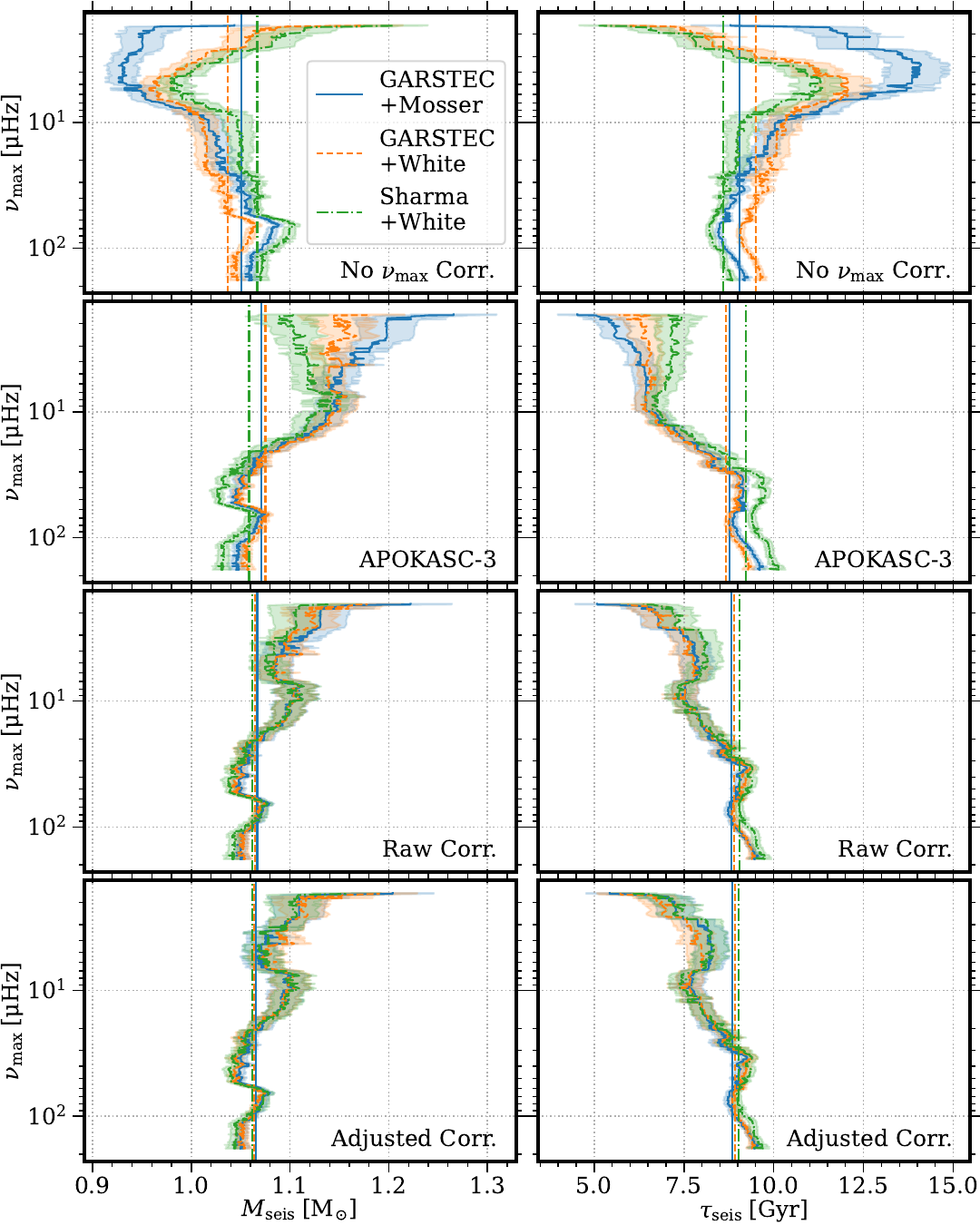}
    \caption{\label{fig:numax_mass_age}Consistency between lower RGB and upper RGB within the $\alpha$-rich population. The two columns study the consistency of asteroseismic masses and ages, respectively. The four rows correspond to four correction schemes; from top to bottom: APOKASC-3 without $\nu_{\rm max}$ correction, APOKASC-3 correction, raw $f_R$ correction, and adjusted $f_R$ correction. In each panel, we compare three combinations of models and weighting approaches: GARSTEC+Mosser (blue solid curves), GARSTEC+White (orange dashed curves), and Sharma+White (green dash-dotted curves). Rolling medians of masses and ages are shown as functions of $\nu_{\rm max}$; horizontal error bars (shaded regions) are median absolute deviations (MADs) multiplied by $k = 1/(\Phi^{-1}(3/4)) \approx 1.4826$, which converts MADs to standard deviations for normal distributions \NewEdit{($\Phi$ is the cumulative distribution function)}; overall medians are shown as vertical straight lines. For GARSTEC+Mosser, median masses and ages are tabulated in Table~\ref{tab:arich_medians}.}
\end{figure*}

Fig.~\ref{fig:numax_mass_age} checks the consistency between lower RGB and upper RGB stars within the $\alpha$-rich population in terms of both asteroseismic masses and ages. To better illustrate the original APOKASC-3 correction scheme, we also visualize the case where no $\nu_{\rm max}$ correction is applied. Specifically, we divide tabulated $M_{\rm seis}$ values by $f_{\nu_{\rm max}}^3$ (instead of $f_{\nu_{\rm max}}$) and adjust $f_{\Delta\nu}$ using the $f_{\Delta\nu}$ nu-SVR described above. As seen in the top row, using the \citet{Mosser2012A&A} weighting approach, without any $\nu_{\rm max}$ correction, upper RGB stars seem to have smaller masses and larger ages than lower RGB stars. The \citet{White2011ApJ} weighting approach yields substantially different masses and ages for luminous giants, regardless of whether GARSTEC or \citet{Sharma2016ApJ} models are used. By calibrating asteroseismic radii to the Gaia-based reference system, APOKASC-3 (shown in the second row) successfully addressed this inconsistency up to $\nu_{\rm max} \sim 7 \,{\rm \mu Hz}$, significant improvement over the no $\nu_{\rm max}$ correction case. However, with Eq.~(\ref{eq:scalingM}), APOKASC-3 seems to have over-calibrated the consistency between lower RGB and upper RGB and flipped the signs for both masses and ages.

With Eq.~(\ref{eq:scalingM_fR}), our raw $f_R$ correction (third row) further reduces the systematics involved in the selection of models and weighting approaches, from $\sim 2\sigma$ to $< 1\sigma$ for luminous giants. Meanwhile, it also significantly reduces the inconsistency between lower RGB and upper RGB. By invoking the feedback loop between $M_{\rm seis}$ and $f_{\Delta\nu}$, our adjusted $f_R$ correction (bottom row) enhances the agreement among GARSTEC+Mosser, GARSTEC+White, and Sharma+White to a level significantly below random errors, making asteroseismic mass measurements much more robust than other correction scheme. The improvement compared to the raw $f_R$ correction scheme is less remarkable in terms of ages, since stellar age is highly sensitive to mass. Regarding the consistency within the $\alpha$-rich population, our adjusted $f_R$ correction outperforms the raw version by $\sim 1\sigma$ on the upper RGB.

\begin{table}
    \caption{\label{tab:arich_medians}Median asteroseismic masses (in ${\rm M}_{\odot}$) and ages (in ${\rm Gyr}$) of the $\alpha$-rich population according to GARSTEC models and the \citet{Mosser2012A&A} weighting approach. We consider four different correction schemes: APOKASC-3 without $f_{\nu_{\rm max}}$ correction, APOKASC-3, raw $f_R$ correction, and adjusted $f_R$ correction. For each combination of stellar property and correction scheme, four quantities are provided: median of all stars, median of lower RGB ($\nu_{\rm max} > 30 \,{\rm \mu Hz}$) stars, median of upper RGB stars ($\nu_{\rm max} < 30 \,{\rm \mu Hz}$), and fractional discrepancy, defined as $({\rm med}[{\rm Upper}] - {\rm med}[{\rm Lower}]) / {\rm med}[{\rm Overall}]$.}
    \centering
    \begin{tabular}{lcccc}
    \hline
        Property (Scheme) & Overall & Lower & Upper & Discrep. \\
    \hline
        Mass (No $f_{\nu_{\rm max}}$) & $1.0502$ & $1.0652$ & $1.0055$ & $-0.0568$ \\
        Mass (APOKASC-3) & $1.0702$ & $1.0533$ & $1.1244$ & $0.0665$ \\
        Mass (Raw) & $1.0664$ & $1.0571$ & $1.0863$ & $0.0274$ \\
        Mass (Adjusted) & $1.0645$ & $1.0574$ & $1.0757$ & $0.0172$ \\
    \hline
        Age (No $f_{\nu_{\rm max}}$) & $9.0576$ & $8.7868$ & $10.5463$ & $0.1943$ \\
        Age (APOKASC-3) & $8.7809$ & $9.1701$ & $7.2548$ & $-0.2181$ \\
        Age (Raw) & $8.8202$ & $9.0481$ & $7.9796$ & $-0.1211$ \\
        Age (Adjusted) & $8.8405$ & $9.0419$ & $8.1978$ & $-0.0955$ \\
    \hline
    \end{tabular}
\end{table}

To better quantify the level of consistency, we further divide the $\alpha$-rich population into two subsamples: $630$ lower RGB stars with $\nu_{\rm max} > 30 \,{\rm \mu Hz}$ and $513$ upper RGB stars with $\nu_{\rm max} < 30 \,{\rm \mu Hz}$. As seen in the lower panel of Fig.~\ref{fig:fdnu_fm_numax}, our $f_R$ correction schemes only cause sub-percent level changes in asteroseismic masses compared to APOKASC-3 values to the lower RGB subsample, making it a reasonable benchmark for testing the mass scale for luminous giants. In Table~\ref{tab:arich_medians}, we tabulate median asteroseismic masses and ages of the entire $\alpha$-rich population and both subsamples based on GARSTEC+Mosser values; furthermore, we define the fractional discrepancy as $({\rm med}[{\rm Upper}] - {\rm med}[{\rm Lower}]) / {\rm med}[{\rm Overall}]$. Compared to APOKASC-3, our adjusted correction scheme reduces fractional discrepancies in median masses and ages from $6.65\%$ to $1.72\%$ and from $-21.81\%$ to $-9.55\%$, respectively, which are significant improvements. The remaining inconsistency may be caused by subtle differences in metallicity distributions between the two subsamples (with \NewEdit{upper RGB} stars having slightly lower metallicities) or temperature-dependent systematics in spectroscopic measurements. We note that potential contamination of AGB stars would tend to assign smaller masses to stars at large radius (small $\nu_{\rm max}$), which is counter to the observed trend in our $f_R$ correction regime.

\begin{figure*}
    \centering
    \includegraphics[width=0.9\textwidth]{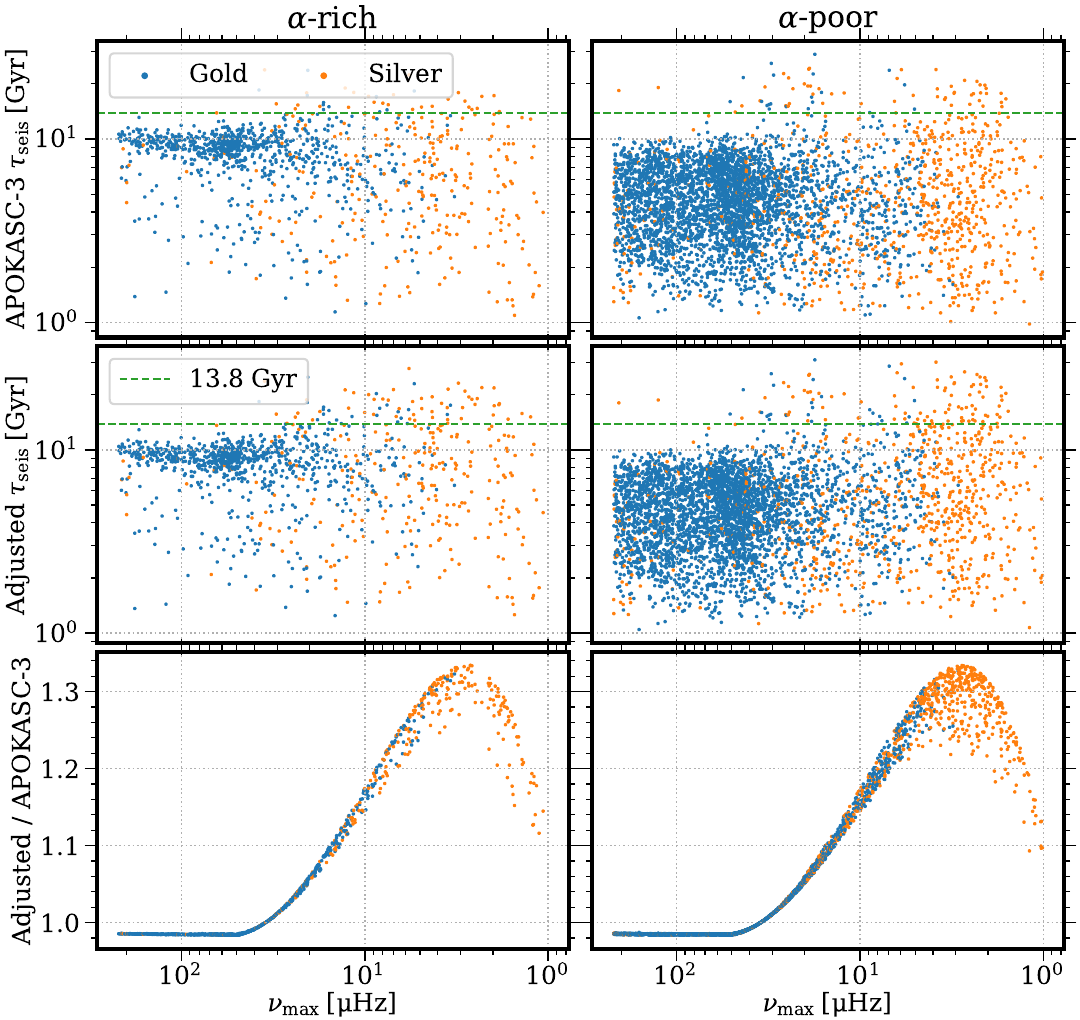}
    \caption{\label{fig:tau_numax}Scatter plots of asteroseismic age $\tau_{\rm seis}$ versus frequency of the maximum power $\nu_{\rm max}$. The $4439$ ``Gold'' sample stars are shown in blue, while the $1097$ ``Silver'' sample stars are shown in orange. The two columns present the $\alpha$-rich and $\alpha$-poor populations\NewEdit{. The first} two rows correspond to ages from APOKASC-3 \citep{Pinsonneault2025ApJS} and our adjusted $f_R$ correction, respectively\NewEdit{, and the last row shows the ratios between these two sets of age predictions}. \NewEdit{The age of the Universe ($\sim 13.8 \,{\rm Gyr}$) is shown as green dashed horizontal lines in the first two rows.} All ages shown in this figure are based on GARSTEC \citep{Weiss2008Ap&SS} models and the \citet{Mosser2012A&A} weighting approach.}
\end{figure*}

In Fig.~\ref{fig:tau_numax}, we investigate the impact of our adjusted $f_R$ correction scheme on predicted ages for both $\alpha$-rich and $\alpha$-poor populations. \NewEdit{It is clear} that lower RGB (high $\nu_{\rm max}$) stars are barely affected, while upper RGB (low $\nu_{\rm max}$) stars are predicted to be older as they are assigned lower masses by $f_R$ correction. A small fraction of stars exceed the age of the Universe ($\sim 13.8 \,{\rm Gyr}$). This can be explained by large uncertainties in mass or mechanisms like binary mass transfer.

\begin{figure}
    \centering
    \includegraphics[width=\columnwidth]{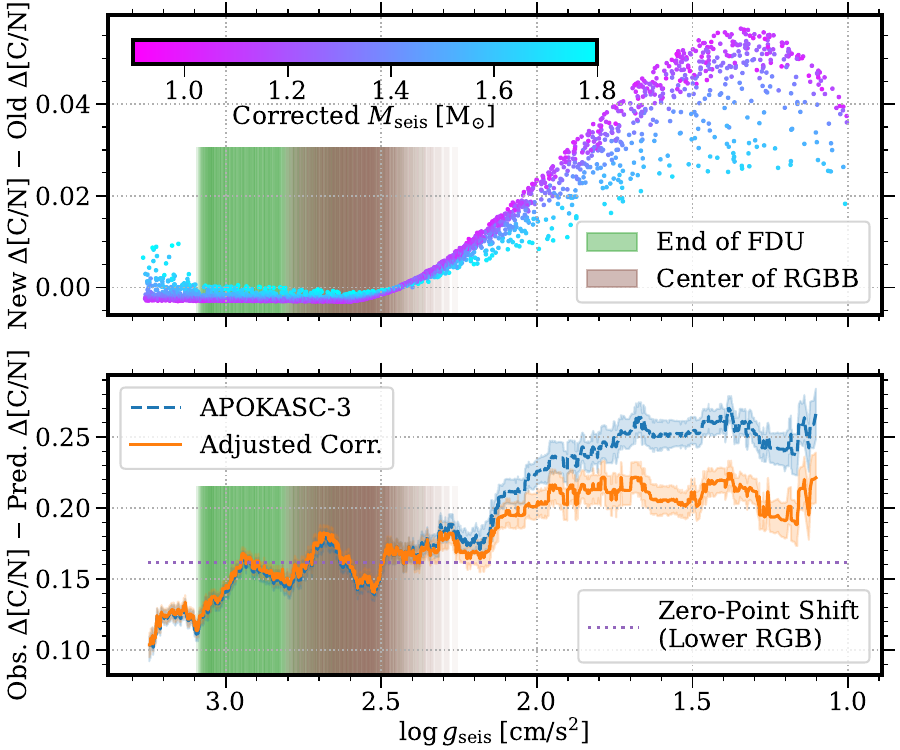}
    \caption{\label{fig:dcn_logg}Impact on first dredge-up predictions for the $\alpha$-poor population. Each panel plots the difference between two versions of surface carbon-to-nitrogen abundance ratio changes $\Delta[{\rm C}/{\rm N}]$ (current minus birth) as a function of logarithmic seismic gravity. Upper panel: difference between post-FDU $\Delta[{\rm C}/{\rm N}]$ values predicted for corrected (adjusted $f_R$ correction; ``New'') and APOKASC-3 (``Old'') asteroseismic masses. Note that the color-coding has been switched to corrected $M_{\rm seis}$. Lower panel: rolling medians of discrepancies between observed \citep[with birth trends removed following][]{Roberts2024MNRAS} and predicted $\Delta[{\rm C}/{\rm N}]$; vertical error bars (shaded regions) are MADs multiplied by $k \approx 1.4826$. The median trends using APOKASC-3 and corrected masses are shown as blue dashed and orange solid curves, respectively. In both panels, the end of FDU and the center of red giant branch bump are shown as green and brown bands, respectively. For both quantities, we use benchmark predictions from \citet{Cao2025ApJ}; for the latter, we superimpose an observed zero-point shift of $0.1509 \,{\rm dex}$ \citep[see Section~5.2 of][]{Cao2025ApJ}.}
\end{figure}

The first dredge-up (FDU), changes in surface chemical composition of lower RGB stars, is of particular interest to galaxy archaeologists. Its main diagnostic, the surface carbon-to-nitrogen abundance ratio $[{\rm C}/{\rm N}]$, is sensitive to stellar mass and can be used as an age proxy \citep[e.g.,][Roberts et al. 2025, \NewEdit{submitted}]{Martig2016MNRAS, Ness2016ApJ}. \NewEdit{In the upper panel of Figure~\ref{fig:dcn_logg}, we present the changes in predicted $\Delta[{\rm C}/{\rm N}]$ caused by the mass correction. As upper RGB stars are assigned lower masses by our new correction scheme, they are predicted to have less negative post-FDU $[{\rm C}/{\rm N}]$ values. While the differences are significant, they are relatively small compared to the discrepancy between data and theory \citep{Cao2025ApJ}.}

\NewEdit{However, it is interesting to study the gravity dependence of such discrepancy.} In \citet{Cao2025ApJ} Figure~\NewEdit{3,\footnote{Figure~5 in the initial submission (first version on arXiv).}} we found that the discrepancy between observed and predicted $\Delta[{\rm C}/{\rm N}]$ increases \NewEdit{as stars} ascend the red giant branch. \NewEdit{As we can see from the lower panel of Figure~\ref{fig:dcn_logg}, the inconsistency between lower RGB and upper RGB is again significantly reduced by our adjusted $f_R$ correction scheme. We note that extra mixing \citep{Shetrone2019ApJ} in luminous giants would cause $[{\rm C}/{\rm N}]$ to decrease as stars ascend the red giant branch and thus could not explain the increase we see here. Besides, the gravity trend at the bottom of the RGB (leftmost portion of the plot) is due to incomplete FDU (Liagre et al. 2025, submitted) in theoretical models, a feature not seen in APOKASC-3 \citep{Cao2025ApJ}.}

\section{Summary} \label{sec:summ}

Accurately determining asteroseismic masses for luminous giants is challenging due to systematics in scaling relations. In APOKASC-3 \citep{Pinsonneault2025ApJS}, we attempted to address this issue by correcting $\Delta\nu$ and ${\nu_{\rm max}}$ using theoretical models and astrometric radii, respectively. In this work, we have proposed and investigated an alternative formulation of the correction scheme. We have demonstrated that our adjusted correction substantially reduces the systematics concerning the choice of models \citep{Weiss2008Ap&SS, Sharma2016ApJ} and weighting approaches \citep{White2011ApJ, Mosser2012A&A}. Meanwhile, it improves consistency between lower RGB and upper RGB in several aspects, most importantly asteroseismic age of the $\alpha$-rich population and first dredge-up signals of the $\alpha$-poor population. We caution that discrepancies between different correction schemes should be included as an integral part of asteroseismic systematics in either case.

To conclude, we notice that the advent of APOGEE Data Release 19 \citep{Meszaros2025AJ} will provide important information about spectroscopic systematics, which we will further investigate in the future. Meanwhile, with stellar oscillations better understood, e.g., with the forthcoming Nancy Grace Roman Space Telescope \citep{Weiss2025ApJ}, we should be able to obtain more accurate asteroseismic masses for luminous giants.

\section*{Acknowledgments}

We are grateful for judicious feedback from Amanda Ash, Jamie Tayar, Jack Roberts, Joel C. Zinn\NewEdit{, and the anonymous referee}. We acknowledge insightful comments from other colleagues at OSU, especially Jennifer A. Johnson and Lucy Lu. KC acknowledges support from NASA grant 22-ROMAN11-0011. MHP is supported by NASA grant 80NSSC24K0637.

\software{{\sc pandas} \citep{Reback2022zndo}, {\sc NumPy} \citep{Harris2020Natur}, {\sc SciPy} \citep{Virtanen2020NatMe}, {\sc scikit-learn} \citep{Pedregosa2011JMLR}, {\sc Matplotlib} \citep{Hunter2007CSE}}

\section*{Data Availability}

The APOKASC-3 data catalog is available in the published paper \citep{Pinsonneault2025ApJS}. \NewEdit{\citet{Cao2025ApJ} details the simulation results used in Section~\ref{sec:res} and shown in Figure~\ref{fig:dcn_logg}. Those MESA inlists are available in Zenodo under an open-source Creative Commons Attribution license at doi: \url{https://doi.org/10.5281/zenodo.15863879} \citep{10.5281/zenodo.15863879}. The} simulation results are available on request.

\bibliography{main}{}
\bibliographystyle{aasjournal}

\end{document}